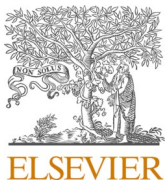
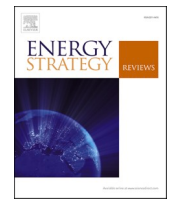

# MENA compared to Europe: The influence of land use, nuclear power, and transmission expansion on renewable electricity system costs

H. Ek Fälth [a,*], D. Atsmon [a], L. Reichenberg [a,b], V. Verendel [c]

[a] *Department of Space Earth and Environment, Chalmers University of Technology, 412 96, Göteborg, Sweden*
[b] *Department of Mathematics and Systems Analysis, Aalto University, Otakaari 1 F, Espoo, Finland*
[c] *Department of Computer Science and Engineering, Chalmers University of Technology, 412 96, Göteborg, Sweden*

ABSTRACT

Most studies that examine $CO_2$-neutral, or near $CO_2$-neutral, power systems by using energy system models investigate Europe or the United States, while similar studies for other regions are rare. In this paper, we focus on the Middle East and North Africa (MENA), where weather conditions, especially for solar, differ substantially from those in Europe. We use a green-field linear capacity expansion model with over-night investment to assess the effect on the system cost of (i) limiting/expanding the amount of land available for wind and solar farms, (ii) allowing for nuclear power and (iii) disallowing for international transmission. The assessment is done under three different cost regimes for solar PV and battery storage.

First, we find that the amount of available land for wind and solar farms can have a significant impact on the system cost, with a cost increase of 0–50% as a result of reduced available land. In MENA, the impact on system cost from land availability is contingent on the PV and battery cost regime, while in Europe it is not. Second, allowing for nuclear power has a minor effect in MENA, while it may decrease the system cost in Europe by up to 20%. In Europe, the effect on system cost from allowing for nuclear power is highly dependent on the PV and battery cost regime. Third, disallowing for international transmission increases the system cost by up to 25% in both Europe and MENA, and the cost increase depends on the cost regime for PV and batteries.

The impacts on system cost from these three controversial and policy-relevant factors in a decarbonized power system thus play out differently, depending on (i) the region and (ii) uncertain future investment costs for solar PV and storage. We conclude that a renewable power system in MENA is likely to be less costly than one in Europe, irrespective of future uncertainties regarding investment cost for PV and batteries, and policies surrounding nuclear power, transmission, and land available for wind- and solar farms. In MENA, the system cost varies between 42 and 96 $/MWh. In Europe, the system cost varies between 51 and 102 $/MWh.

## 1. Introduction

The 2015 UN climate summit in Paris (COP21) demonstrated a broad consensus on the need for comprehensive action to reduce greenhouse gas emissions and keep global warming in check. The electricity sector is a major contributor to $CO_2$ emissions, accounting for around one-quarter of total emissions [1]. Global electricity consumption is projected to grow due to improved living standards in developing economies [2] and electrification of other sectors, such as transportation [3]. Meanwhile, mitigating $CO_2$ emissions in the power sector is less expensive than in other sectors [3]. For these reasons, the literature on $CO_2$-neutral,[1] or near-$CO_2$-neutral, power systems is expanding. A subset of the literature on $CO_2$-neutral power systems uses investment models to investigate economic feasibility for different conceivable future power systems. Many of these studies span entire continents and employ a large share of variable renewable energy (VRE), i.e. they are systems that mainly rely on solar irradiation and wind [4–29]. The majority model Europe, the United States, or other temperate regions, while continents with warmer climates have received less attention. There are, however, a few such studies: Aghahosseini et al. studied the Middle East and North Africa region (MENA) [15], Barbosa et al. studied South and Central America [10] and, Blakers et al. studied Australia

---

* Corresponding author.
  *E-mail address:* hannafa@chalmers.se (H.E. Fälth).
[1] By the term $CO_2$-neutral power systems, this study refers to a system consisting of generation technologies that do not emit $CO_2$ during power generation, such as solar-, wind-, and nuclear power. Biogas is also considered CO2-neutral, despite its controversies.

[30]. With the exception of [15], MENA has been modeled mainly as a potential provider of solar power for Europe [13,31,32]. However, the MENA region merits investigation in its own right, not least because of its current reliance on fossil fuels, with a power plant mix comprising 68% natural gas and 23% oil [33]. The high carbon intensity of the MENA electricity generation, improving living standards in the region, concerns about pollution, and the possibility of electrification of, for instance, transportation, entail large potential benefits of decarbonizing the MENA power sector. Aghahosseini et al. show that a 100% renewable energy system in MENA in 2030 can be less costly than the system corresponding to a BAU trajectory. Thus, both the weather conditions, which are different relative to those of more commonly studied power systems, particularly in terms of more abundant solar resources, and the urgent need to replace carbon-intensive power generation, motivate giving MENA more attention from energy-system studies.

The prospect of $CO_2$-neutral power systems raises different public concerns, and the literature on social acceptance of renewable energy technologies and associated infrastructures has expanded [34]. Large-scale wind and solar farms, nuclear power, and transmission expansion are three issues commonly addressed in this literature [35–49]. Bolwig et al. [43] argue that social acceptance is important to consider in energy system modeling, policy and planning due to its impact on consumer costs, energy mix, and revenue distribution.

Regarding wind- and solar power deployment, large-scale farms have sparked local resistance and are of public concern [37] [–] [39,50,51]. With increased competition of land and environmental concerns for wind and solar power, suitable sites for wind and solar farms could be a constraint necessary to consider in energy system planning. A limited access to suitable sites for wind- and solar power could have an effect on system cost for large scale $CO_2$-neutral power system. However, potential constraints due to public concern regarding large-scale wind- and solar farms have received little attention in the energy system modeling community. Schlachtberger et al. [4] found an increase in system cost in Europe by about 10% when the land available for onshore wind was reduced to zero. Bolwig et al. [43] modeled the Nordic-Baltic region and found that the consumer costs for electricity could increase with 12% as a result of low social acceptance for wind power. Both of these studies were done on temperate regions, and the investigation regarding the availability of land focused on wind-, rather than both wind- and solar-, power.

The role of nuclear power has been investigated in several energy system studies [17,52–57], although only in temperate regions and with a large spread regarding its resulting potential to reduce system cost. Nuclear power is a contentious issue, both in society at large [45–49] and in the modeling community [52,53]. Some authors have argued that nuclear power (or other carbon-neutral baseload technologies) is a crucial technology for keeping costs down in a future low carbon emissions power system [17,54,55], while others find only moderate cost benefits of including nuclear power [56,57]. Yet other studies exclude nuclear power by design and find that a future power system based on VRE may be achieved at low to moderate cost [15,16,18]. Sepulveda et al. [17] model systems with and without what they term "firm low-carbon technologies" (essentially CCS technologies and nuclear), and find that excluding such technologies increases the system cost by 10–100%. Jägemann et al. [56] found a cost difference of between 11 and 44% between a system with or without nuclear in Europe, depending on the investment costs mainly for wind and solar. Pattupara et al. [57] modeled Switzerland and neighboring countries and found a 15% decrease in system cost when nuclear power was allowed, while Hong et al. [55] found that replacing current nuclear power in Sweden with wind- and solar power would yield a system cost around five times higher than the current cost for electricity. However, these previous studies all apply different system boundaries and differ both with regards to trading of variations through transmission, as well as the inclusion of long-term storage options.

Transmission expansion has shown to be essential to keep costs down in electricity systems dominated by VRE [4,6–15,58,59], with previous studies showing a cost decrease of about 10–30% if continental grid connections is allowed [6,7,10,11,13,14,58,59]. However, massive transmission expansion may not be politically feasible or publicly acceptable [35,40–44]. The transmission expansion in the EU is slow, despite promotion from the EU commission, with the critical reasons being regulatory issues and permitting issues including bureaucracy and public opposition [60,61].

This paper investigates the importance of these three controversial issues associated with $CO_2$-neutral power systems for two regions with different weather conditions, Europe and MENA. We test the effects on system cost of (i) different levels of restriction on land use for wind- and solar deployment, (ii) allowing/not allowing nuclear power, (iii) allowing/not allowing international transmission. In addition, we examine how conditions that may be known from readily available data, such as population density, land area, and climate, may be used to predict the cost of a renewable power system. We use MENA and Europe as test cases since they differ in terms of resources. Applying the model to Europe allows us to benchmark our results against those in the literature, e.g., Ref. [4,11,24]. The overarching research questions in this paper are:

1. What is the cost of a $CO_2$-neutral future power system in MENA/Europe?
2. What is the impact of weather conditions and demand density on the cost of $CO_2$-neutral power systems?
3. What is the impact of (i)-(iii) on system cost?

Hence, this study contributes to the literature gap in two main areas regarding $CO_2$-neutral power systems. Firstly, by investigating how different climates can affect system costs. Secondly, by examining how the three aforementioned socio-political factors can influence economic feasibility. The paper is organized as follows: Section 2 describes the scenarios, model, and data input, and provides resource availability in the two regions in the form of supply curves for wind and solar. In Section 3, the results for the four scenarios (base, land availability, nuclear, and transmission expansion) are presented, and we discuss the results relative to the literature as well as policy. Section 4 provides concluding remarks and indicates a direction for future research.

## 2. Method

We develop a green-field linear capacity expansion model with overnight investment to model a future power system in MENA and Europe. The model minimizes the total cost for a $CO_2$-neutral power system that meets the electricity demand in each region at each hour in 2040. The model is described in detail in Section 2.1. By evaluating both regions, MENA and Europe, using the same model, the difference in results between the two regions may be attributed to differences in demand and weather, rather than different model formulations and cost assumptions. Hence, we can examine the first and the second research question: the comparison between MENA and Europe regarding the system cost of a $CO_2$-neutral power system in MENA and Europe and the impact of weather conditions and demand density on these costs.

In order to examine the third research question, the effects on system cost of (i) different levels of restriction on land that may be used for wind-and solar deployment, (ii) allowing/not allowing nuclear power, (iii) allowing/not allowing international transmission, four different scenarios are evaluated: One base scenario as a reference, and three scenarios where the conditions regarding transmission expansion, nuclear power, and land availability for wind- and solar farms are varied, see Table 1. Nuclear power and international/inter-subregional transmission are either available as investment options without any upper limit on capacity, or they are excluded by removing that technology as an investment option. As an upper limit on possible investments for solar and wind power capacity, all scenarios restrict land available for wind-and solar exploitation by excluding unsuitable locations for wind- and solar farms (see section 2.2.3 for more details). We then assume that a percentage of the remaining land is available for onshore wind- and utility solar PV. This percentage is varied between 2 and 20%, in order to examine the effect on system cost from different levels of restriction on land use for wind- and solar deployment. The percentages are indicated in Table 1 and apply to each technology, e.g. 10% means that 10% of the remaining land is available for wind power and 10% of the remaining land is available for solar power, which makes 20% of the remaining land available for onshore wind- and utility solar PV in total.

**Table 1**
Modeled scenarios.

| Scenario | Nuclear Power | Transmission | Available land [%of remaining land] |
| --- | --- | --- | --- |
| Base | No | Yes | 10 |
| Varying land restriction | No | Yes | 2–20 |
| Nuclear | Yes | Yes | 10 |
| No Transmission | No | No | 10 |

All scenarios are evaluated for high-, mid- and low PV- and battery costs, see Table 2. The PV costs are retrieved as the low, mid, and high cost-scenario projections by NREL [62]. In addition to the technologies listed in Table 2, there is the option of residential PV, PV rooftop. The cost for PV rooftop is assumed to be 50% higher than the cost for PV Utility, see Section 2.2.2 below. The evaluated costs for batteries are retrieved from utility-scale lithium-ion storage projections made by Cole [63], as the highest, midrange, and lowest projected costs.

### 2.1. Model

We develop a linear capacity expansion model with hourly resolution for a full chronological year, to minimize total system cost for a power system that meets the demand at all times. Since the focus is to evaluate the cost-efficiency of a future system with inter-continental grid connections, rather than the pathway to reach such a system, we employ over-night investment in a green-field optimization approach. The exception is hydropower, which is assumed to be installed at its present capacity, as reported by the World Energy Council [64]. Technology costs and electricity demand are projections for 2040. Demand- and weather data, as well as costs and technology performances, are exogenous to the model. The model is implemented in Julia using JuMP, a domain-specific modeling language for mathematical optimization embedded in Julia. Subregion data, capacity factors, capacity limits for solar- and wind power, and electricity demand profiles are generated by the GlobalEnergyGIS package described in Refs. [65] and publicly available at GitHub [66].

Variables subject to optimization are capacity investments, electricity generation, storage, and transmission. These variables are functions of the subregions $R = \{r_1,..r_n\}$, the technologies possible to invest in $K = \{k_1,..k_n\}$, different classes of solar- and wind power $C = \{c_1,..c_n\}$ (depending on capacity factor) and the hours over one year $H = \{h_1,..h_n\}$. Parameters given to the model include technology costs, technology efficiencies, demand, the distance between subregions, and capacity factors. The model represents wind and solar power using five resource classes for each region, see details in the supplementary material. We use the GlobalEnergyGIS package [65,66] to generate the maximum potential capacity (in $GW$) and hourly capacity factors for each technology, resource class, and region.

**Table 2**
PV- and battery costs.

|  | High-Costs | Mid-Costs | Low-Costs |
| --- | --- | --- | --- |
| PV Utility [$/kW] | 1200 | 800 | 400 |
| Battery [$/kWh] | 375 | 230 | 87.5 |

The objective function to be minimized is the total system cost. The total system cost ($SC$, [$M€/year$]) is a function of electricity generation ($G_{r,k,c(k),h}$, [$GWh/h$]), operation and management cost ($omc_k$, [$€/GWh$]), fuel cost ($fuc_k$, [$€/GWh$]), technology efficiency ($\eta_k$, [ − ]), installed capacity ($C_{r,k,c(k)}$, [$GW$]), investment cost ($ic_k$, [$€/GW$]), annualisation factor ($af_k$), fixed cost ($fc_k$, [$€/GW/year$]), transmission capacity ($TC_{r_1,r_2}$, [$GW$]) and transmission cost ($tc_{r_1,r_2}$, [$€/GW$]).

$$SC = \sum_{r,k,c(k),h} G_{r,k,c(k),h}(omc_k + fuc_k / \eta_k) + \sum_{r,k,c(k)} C_{r,k,c(k)}(ic_k \cdot af_k + fc_k) + 0.5\sum_{r_1,r_2} TC_{r_1,r_2} \cdot tc_{r_1,r_2} \quad (1)$$

By convention we use uppercase for variables and lowercase for parameters.

The transmission cost is divided by two since the model is investing in transmission lines between subregions $r_1$ and $r_2$ and between $r_2$ and $r_1$, even though only one line is needed. The transmission cost (tc, [$€/GW$]) is a function of transmission line cost ($tlc$,[$€/GW/km$]), distance between subregions ($di_{r_1,r_2}$, [$km$]), transmission substation cost ($tsc$, [$€/GW$]) and a fixed transmission cost ($tfc$, [$%ofic$]). Two substations are assumed to be needed for each transmission line.

$$tc_{r_1,r_2} = (tlc \cdot di_{r_1,r_2} + 2 \cdot tsc) \cdot (af + tfc) \quad (2)$$

The implemented constraints follow. First, we need to ensure load balance, i.e., make sure that demand is met at all times. The total generated electricity ($G$, [$GWh/h$]) less the electricity used for storage charging ($CH$, [$GWh/h$]) plus the imported electricity ($TG_{r_2,r,h}$, [$GWh/h$]) and less the exported electricity ($TG_{r,r_2,h}$, [$GWh/h$]) must be greater than or equal to the demand ($d$, [$GWh/h$]), for each hour in every subregion. The transmission losses ($tl$, [$\%/1000km$]) depends on the distance between subregions ($di_{r_1,r_2}$, [$km$]) and are assumed to occur only for imports, to avoid double-counting.

$$\sum_{k,c} G_{r,k,c(k),h} - \sum_{k=storage} CH_{r,k,h} + \sum_{r_2} (1 - tl_{r_2,r}) \cdot TG_{r_2,r,h} - TG_{r,r_2,h} \geq d_{r,h} \quad (3)$$

The electricity generation per hour ($G$, [$GWh/h$]) must be less then or equal to the installed capacity ($C$, [$GW$]) multiplied by the capacity factor ($cf$) for each technology, hour and subregion.

$$G_{r,k,c(k),h} \leq C_{r,k,c(k)} \cdot cf_{r,k,c(k),h} \quad (4)$$

The storage level ($SL_{r,k,h}$, [$GWh/h$]) cannot be negative:

$$SL_{r,k,h} \geq 0 \quad (5)$$

The maximum storage level depends on the installed capacity ($C_{r,k,c}$, [$GW$]) and the discharge time ($dt_{r,k}$, [$h$]) for the storage technology. For batteries, the discharge time is set to 8 h; for hydro dams, it depends on the dam size.

$$SL_{r,k,h} \leq C_{r,k,c(k)} \cdot dt_{r,k} \quad (6)$$

The present storage level ($SL_{r,k,h}$, [$GWh/h$]) depends on battery charging ($CH_{r,k,h}$, [$GWh/h$]), the water flow in-to the dams, i.e. a capacity factor for hydro inflow ($cfh_{r,h}$, [ − ]), the installed capacity of hydro dams ($C_{r,dam}$, [$GW$]), the electricity going from the storage to the grid ($G_{r,k,c,h}$, [$GW$]) and the electricity losses. These losses depend on the efficiency for each storage technology ($\eta_k$, [ − ]). The storage balance is written as, for h¿1:

$$SL_{r,k,h} \leq SL_{r,k,h-1} + CH_{r,k,h} + cfh_{r,h} \cdot C_{r,dam} - \frac{G_{r,k,c(k),h}}{\eta_k} \quad (7)$$

If h = 1, the first term after the inequality sign is instead the storage level in the last hour of the previous year. Note that the first term is less than or equal to and not simply equal to. This is due to spillage when the water inflow is greater than the amount of water that the dam can handle.

Charging batteries requires batteries:

$$CH_{r,battery,h} \leq C_{r,battery} \quad (8)$$

Transmission constraints assure that the transmitted electricity ($TG_{r_1,r_2}$, [$GWh/h$]) does not exceed the installed transmission capacity ([$TC_{r_1,r_1}$, $GW$]) and that the installed transmission between subregion $r_1$ and $r_2$ is the same as between $r_2$ and $r_1$.

$$\begin{aligned} TG_{r_1,r_2,h} &\leq TC_{r_1,r_2} \\ TC_{r_1,r_2} &= TC_{r_2,r_1} \end{aligned} \quad (9)$$

In order to partially mimic realistic constraints on nuclear power plants, ramping constraints and a minimum generation level in percentage of installed capacity are imposed.

$$\begin{aligned} G_{r,nuclear,h} &\leq G_{r,nuclear,h-1} + 0.2 \cdot C_{r,nuclear} \\ G_{r,nuclear,h} &\geq G_{r,nuclear,h-1} - 0.2 \cdot C_{r,nuclear} \\ G_{r,nuclear,h} &\geq 0.6 \cdot C_{r,nuclear} \end{aligned} \quad (10)$$

We assume a limited stock of biogas. No more than 5% of the total annual electricity generation can be produced by biogas turbines. This assumption is based on estimates on waste, agricultural residues, and manure [67–70]. There are indeed other feed-stocks for biogas such as forest residues, but there is also a demand for bio energy from sectors not included in this study (heat and transport).

$$\sum_{r,h} G_{r,biogas,h} \leq \sum_{r,k,c(k),h} G_{r,k,c(k),h} \cdot 0.05 \quad (11)$$

### 2.2. Data

Input data include: definition of subregions; estimates of transmission distances between subregions; cost- and performance data for technologies and fuels; hourly subregional demand for a full chronological year; capacity factors; and capacity limits for solar power, wind power, and hydropower. This section contains information on how these input data were retrieved and implemented in the model.

#### 2.2.1. Regions and transmission

MENA is modeled with 13 subregions and Europe with 10. All subregions are treated as "copper plates" internally, i.e., transmission within each subregion is assumed to be unconstrained. HVDC transmission lines are assumed to be available for investment between neighboring subregions. Data on the countries aggregated to each subregion, maps, possible interconnections, and interconnection distances may be found in the supplementary material.

The assumed transmission costs are presented in Table 3 and are retrieved from ETSAP [71], except for the fixed cost which is taken from NREL [72]. The lifetime of HVDC lines is assumed to be 35 years [73].

#### 2.2.2. Technology- and fuel costs

The power generating technology options are wind power (on- and offshore), PV (utility and rooftop), concentrated solar power (CSP) and biogas turbines (GT). The assumed costs and efficiencies for each technology are shown in Table 4. The costs are retrieved from the National Renewable Energy Laboratory's (NREL) Annual Technology Baseline (ATB) Database 2018 [62]. This database contains technology cost projections for 2040 for a low-, mid-, and high-cost scenario; the assumed costs used in this study are the mid-cost scenario projections in the NREL database [62]. For PV rooftop, the cost is assumed to be 50% higher than PV utility due to higher installation costs for smaller systems, which is in line with the cost projections in NREL database 2018 [62]. The investment cost for batteries is retrieved from utility-scale lithium-ion storage projections by Cole [63]. Lifetime and round-trip efficiency for the batteries are also retrieved from Cole [63]. The modeled batteries are assumed to be lithium-ion battery packs with a discharge time of 8 h, as in Ref. [63], and the cost can be converted from $/kW to $/kWh with a factor 8. All investment costs are annualized using a social discount rate of 5%. Our assumption lies within the 2–7% range used in previous studies [4–6,11,12,14–27,29] and is in line with the recommendation of a discount rate of maximum 5% for energy systems optimization models, proposed by Garcia-Gusano et al. [74].

We assume the cost of biogas to be the average biogas cost in Ref. [17], USD 60 per MWh.

#### 2.2.3. Wind- and solar data

The wind- and solar input data were constructed using the GlobalEnegyGIS package [65,66]. In this package, installation limits for wind and solar power capacity are based on assumptions on typical wind and PV farm densities ($W/m^2$) and available land ($m^2$). Several auxiliary datasets are used to exclude areas less suitable for solar- and wind power which, together with the meteorological data, leads to an estimate of solar- and wind potentials for each region. The datasets include population (GPWv4 [75]), land cover (MODIS [76]), protected areas (WDPA [77]), and topography and bathymetry (ETOPO1 [78]) [66]. After masking out unsuitable locations, a certain fraction of the remaining area is considered available for solar and wind farms, see available land in Table 5. Assumptions on typical wind and PV farm densities ($W/m^2$) are used to convert the resulting available area to potential capacity (GW).

Hourly time series with capacity factors for PV and wind power are constructed using the ECMWF ERA5 [79] database and data from the Global Wind Atlas [80]. The procedure for this is described in detail in Refs. [65,66]. The modeled subregions are divided into pixels ($0.01° x 0.01°$), capturing the different solar and wind conditions with an hourly time resolution. Solar irradiation is used to calculate the annual PV capacity factor profiles assuming fixed-latitude-tilt; wind speed is translated into capacity factors based on a power curve for a typical wind park with Vestas 112 3.075 MW wind turbines [65,66]. The pixels in each subregion are aggregated into five classes, depending on yearly average capacity factors for solar- and wind power, to reduce the computational demand. The pixels within a resource class, in each subregion, are assumed to have the same capacity factor time series (the average of all capacity factor time series in those pixels), see Supplementary Material and [65,66].

Supply curves for PV and wind power, based on the input data retrieved with the GlobalEnergyGIS package [65,66], are shown in Fig. 1. The supply curves display the wind- and solar power potential in relation to electricity demand and reveal the resource prerequisites for a VRE based power system for the two different regions. The numbers are percentages of net demand that can be supplied at different Technology levelized cost of electricity (Technology LCOE) for each technology (wind- and solar power). Technology LCOE ($/MWh) is defined as the net present value of all costs related to producing electricity with that technology per electricity output. Net demand is defined as the total electricity demand minus hydropower generation. The supply curves in Fig. 1 are the results of an assumption of 10% available land for wind and solar deployment (see section 2 for the definition of available land) and midrange costs for PV (see Table 2). It can be seen in Fig. 1 that MENA can produce more of its electricity demand with either wind- or solar power to a lower cost than can Europe. Supply curves for other costs and assumptions on available land are presented in the Supplementary Material.

**Table 3**
Transmission costs.

| Trans. Line [$/MW/km] | Substations [$/MW] | Losses [%/1000 km] | Lifetime [yr] | Fixed O&M Cost [% of Inv. Cost] |
|---|---|---|---|---|
| 2030 | 17,350 | 3 | 35 | 0.8 |

**Table 4**
Technology costs and efficiencies.

|  | Investment Cost [$/kW] | O&M Cost [$/MWh] | Fixed Cost [$/kW/yr] | Lifetime [yr] | Efficiency [−] |
| --- | --- | --- | --- | --- | --- |
| Nuclear | 5570 | 2 | 99 | 60 | 0.32 |
| Wind Onshore | 1227 | 0 | 42 | 25 | – |
| Wind Offshore | 2317 | 0 | 130 | 25 | – |
| PV Utility | 800 | 0 | 6 | 25 | – |
| PV Rooftop | 1200 | 0 | 6 | 25 | – |
| CSP | 5225 | 3.5 | 50 | 30 | – |
| GT | 830 | 4 | 11 | 30 | 0.38 |
| Hydro Power | 0 | 0 | 0 | – | – |
| Battery | 1850 (230 $/kWh) | 1.32 | 6 | 15 | 0.9 |

**Table 5**
Capacity limit assumptions.

|  | PV Utility | PV Rooftop | CSP | Wind Onshore | Wind Offshore |
| --- | --- | --- | --- | --- | --- |
| Density [W/$m^2$] | 45 | 45 | 35 | 5 | 8 |
| Available land [% of remaining land area] | 10 | 10 | 10 | 10 | 33 |

### 2.2.4. Hydropower

Installed hydropower capacities and annual generation in each subregion are assumed to be as in 2016 according to the World Energy Council [64] and can be found in the Supplementary Material. Monthly hydro inflow profiles for each region were retrieved from the GlobalEnergyGIS package [65,66] which uses the GRanD database [81,82]. The inflow profiles are converted to hourly inflow assuming an even flow within each month and the dam size is assumed to be equal to the annual generation divided by 12, i.e., the dam can roughly hold a months's worth of energy before spillage occurs, depending on the inflow profile.

### 2.2.5. Hourly demand profiles

Due to a lack of comprehensive real-world demand data for each of our 23 subregions, we generate synthetic hourly electricity demand using machine learning to algorithmically generate a model directly from available data, and then extrapolate the resulting series to 2050. More specifically, we fit a gradient boosting regression model with decision trees as base learners [83] and evaluate it using cross-validation on data for currently available countries, and then scale the generated time series to fit a projected yearly total in 2050. We model the profile of how the current hourly electricity demand per capita varies over the year as a function of three types of data. First, to describe a country/region, we include purchase-power adjusted regional GDP [84] and gridded global population of the world (GPW-4) [85]. Second, we include information about temperature time series for 2015 from NASA MERRA-2 [86] with country-level temperature averages and extremes, as well as variables for the hour of the day, weekday, and month of the year. Finally, we have country-level annual electricity generation for 2017 from Ref. [87] scaled to match regional final electricity demand in the SSP2-34 scenario for 2050 [84]. After fitting the model based on electricity demand for 44 countries to the normalized demand time series, the resulting predictions for our regions are extrapolated to 2050 by scaling the total to the level corresponding to SSP2-34 scenarios. For more details and evaluation of the approach with gradient boosting regression, see Ref. [65]. The resulting synthetic demand series are treated as inelastic in the optimization model.

## 3. Results and discussion

The results are structured as a comparison between MENA and Europe regarding system cost in general (Section 3.1); when varying land availability (Section 3.2); with the investment option of nuclear power (Section 3.3); and excluding the option of inter-subregional transmission (Section 3.4), see Table 1 for the specifics of each scenario. Each scenario was investigated with a range of cost projections for solar PV and battery storage (see Table 2). The obtained system costs are presented as System levelized cost of electricity (System LCOE), in $/MWh, defined as total electricity system cost per total demand subtracting the annual hydropower generation. Hydropower generation is subtracted since hydropower is modeled as no-cost in this study.

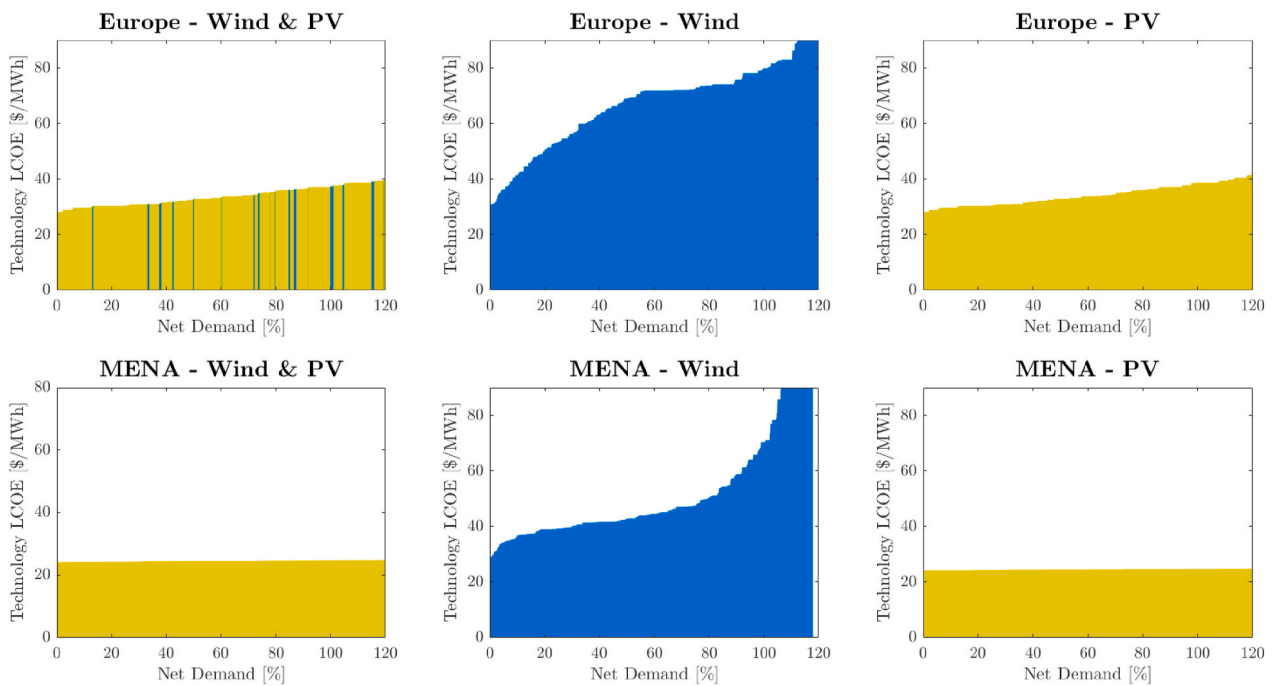

**Fig. 1.** Supply curves for PV and wind power assuming 10% available land and midrange costs for PV.

### 3.1. Comparison between europe and MENA: system cost and technology mix

We begin by presenting a summary of the cost range obtained by modeling the different scenarios as well as PV and battery costs. We find that System LCOE varies substantially, between 51 and 102 $/MWh (Europe) and 42–96 $/MWh (MENA), depending on scenario and PV and battery costs (see Fig. 2). Previous cost estimates in MENA [15] and Europe [4,11,24] lie in this interval. The System LCOE corresponding to midrange costs for PV and batteries for the base scenario, a system with inter-subregional transmission and without nuclear power, is 72 $/MWh for Europe and 61 $/MWh for MENA (denoted with a black line in Fig. 2).

Depending on the scenario, and the cost level for PV and battery investments, the System LCOE is 6–35% lower for MENA compared to Europe. Thus, for any given assumption on land availability, PV and battery costs, transmission expansion, and nuclear power option, the System LCOE is lower in MENA than in Europe.

Fig. 3 shows the optimal generation mixes as a share of total electricity generation in the base scenario for low-, mid- and high-cost PV and batteries for MENA and Europe. All demanded electricity is generated by either hydropower, wind power, PV, or biogas turbines, and together adds up to 100% in Fig. 3. Transmission and batteries are displayed as the shares of total power generation that pass through transmission lines and battery storage, respectively, which is why the total generation exceeds 100% in Fig. 3. The generation mix is dominated by wind power for high- and mid-cost PV and batteries. In these cases, more electricity is traded through transmission lines than is stored and delivered through battery storage. For low-cost PV and batteries, the generation mix is instead dominated by solar power. In this case, more electricity is stored and delivered through battery storage than is traded through transmission lines, especially so in MENA, see Fig. 3. For the optimal generation mix for the other scenarios, see the Supplementary Material.

Our resource quality assessment (Fig. 1) shows that MENA can satisfy its demand with either solar or wind power at a lower Technology LCOE than Europe. We also find that for all investigated scenarios, MENA has a lower System LCOE than Europe. Our results thus show a correlation between resource quality and system cost. However, many additional factors determine the cost and capacity mix of optimal renewable power systems, including the available variation management strategies, e.g. the abundance of reservoir hydropower, as well as the nature of spatial and temporal variations in VRE resources. Solar and wind power display different spatial and temporal variations, with solar power having a diurnal pattern, while wind power generation displays variations on both shorter and longer time scales [88,89]. Complexities associated with how technologies complement each other in time and space influence cost and other features of a renewable power system, including the technology mix. Without prior knowledge of the system properties of wind and solar, respectively, it would, for instance, be difficult to infer that the system mix in the base scenario for midrange costs on PV and batteries is dominated by wind (Fig. 3, middle bar), since solar sites with lower Technology LCOE than wind are abundant in both Europe and MENA (Fig. 1, left column). However, our results still indicate that some factual information on System LCOE may be inferred simply from considering the supply curves. In order to understand that connection better, more research is needed, both on other regions and with other technology scenarios.

### 3.2. Amount of land available for wind and solar farms

By decreasing the land available for wind and solar farms, the supply curves (Fig. 1) become steeper. It is then necessary to exploit sites with a lower output to cover the demand, thus increasing System LCOE. Assuming less land is available for wind and solar farms (2% instead of 10% as in the base scenario) increases the System LCOE by up to 47% in MENA and 25% in Europe, see Fig. 4. In Europe, the increase in System LCOE due to less land being available is up to 23–25%, regardless of PV and battery costs. See Section 2 for the definition of available land applied in this study.

In MENA, the increase of System LCOE due to less available land for VRE exploitation is dependent on investment costs for PV and storage, see Fig. 4, with land availability having almost no effect for low investment costs and a significant effect for high investment costs. This has to do with the technological LCOE for wind power being significantly more affected by available land compared to PV (as may be seen from Fig. 1). This shows also in the resulting generation mix: When there is less available land, deployment of wind decreases, while solar generation increases, an example of which may be seen in Fig. 5.

The reason the System LCOE does not increase in MENA as land becomes more scarce in the low cost case may be explained by the generation mix already being dominated by solar in that case, and thus the impact of the land constraint on wind power is less relevant, see Fig. 6. Solar figures prominently in the mix for Europe too, in the low cost case, but wind is still important, see Fig. 3.

Both wind and solar power face social opposition in some regions in the world [37–39,50,51]. We show that the assumptions on available land are indeed an important determinant for the System LCOE of a $CO_2$-neutral system. In fact, if the available land for VRE exploitation is reduced from 10% to 2%, the System LCOE may increase by up to 25% in Europe and 47% in MENA, depending on the cost for solar PV and battery storage, see Fig. 4. Schlachtberger et al. [4] found an increase in system cost by 10% when the land available for onshore wind was reduced to zero. Unlike in the present study, the set-up used in

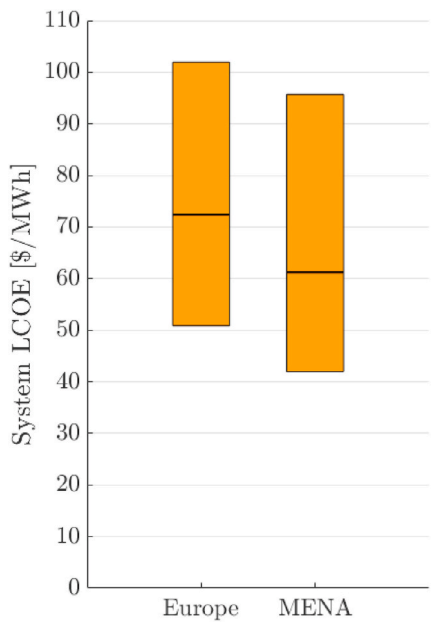

**Fig. 2.** Range of System LCOE in Europe and MENA, obtained by modeling the different scenarios as well as PV and battery costs. The black line corresponds to the System LCOE assuming midrange costs for PV and batteries for the base scenario.

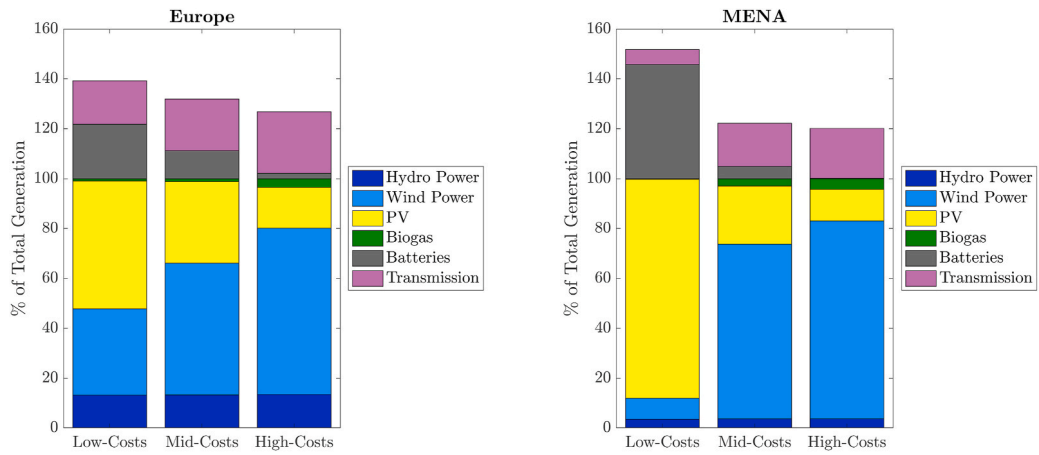

**Fig. 3.** Optimal generation mix in MENA and Europe.

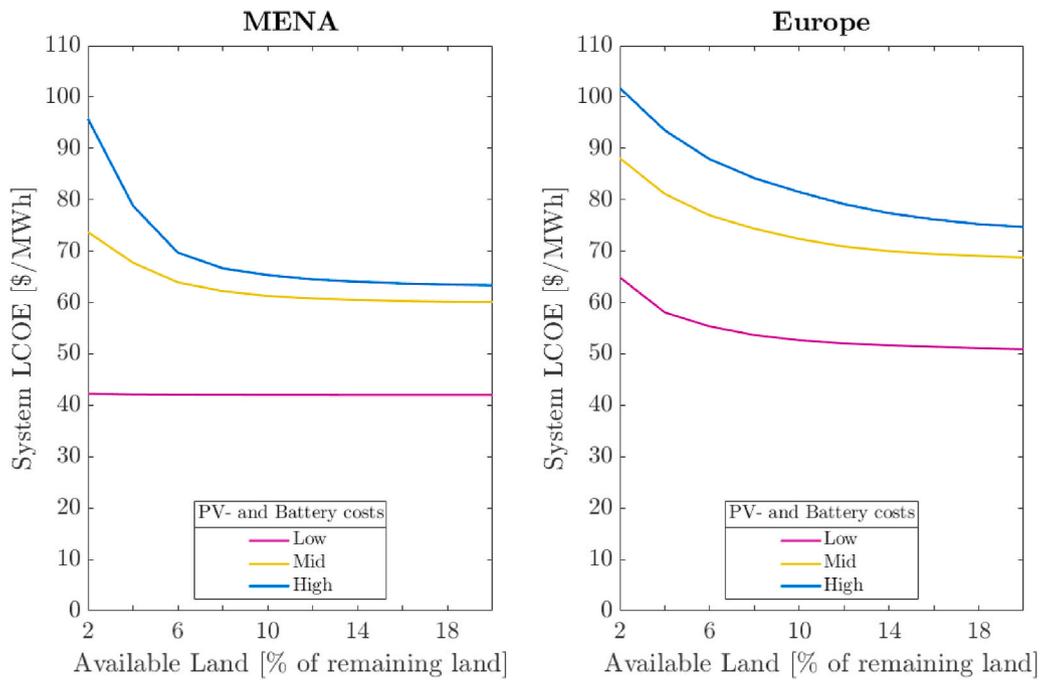

**Fig. 4.** System LCOE as a function of land available for PV and wind power for the cases of low, mid-range and high investment costs for PV and batteries.

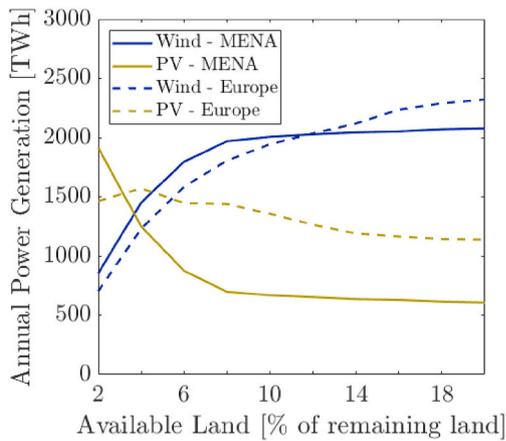

**Fig. 5.** Annual power generation from PV and wind power as a function of available land assuming mid-costs for PV and batteries.

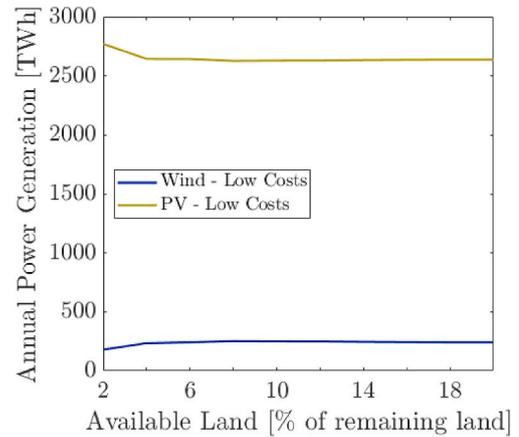

**Fig. 6.** Annual power generation from PV and wind power in MENA as a function of available land assuming low-costs for PV and batteries.

Schlachtberger et al. [4] reduces land available only for on-shore wind power, rather than for both on-shore wind- and solar power. This also applies to Bolwig et al. [43], who examine the effect of 'low social acceptance'[2] on consumer cost in the Nordic-Baltic region by increasing the investment cost for on-shore wind power. They find that the consumer cost for electricity could increase by up to 12% as a result of low social acceptance. They also find that solar PV occupies a larger share of the energy mix if there is low acceptance for wind power. Thus, these previous studies [4,43] have examined the effect of social acceptance for deployment of renewable energy by constraining on-shore wind power, while our study examines the possible effect of social acceptance by constraining both wind- and solar power. Our study shows that the effect of limiting available land for both wind- and solar power in MENA is highly dependent on solar and battery costs, while the effect of constraining available land in Europe depends to a lesser extent on future investment costs. In Europe, the impact on System LCOE of the assumptions on available land are of the same magnitude as the impact of allowing for inter-subregional transmission or nuclear power. In MENA, the effect of available land for wind-and solar farms is more significant than the effect of allowing for inter-subregional transmission or nuclear power. This suggests that land-availability assumptions should feature more prominently than currently in policy discussions and in the modeling community. The difference in System LCOE incurred by assumptions on available land could, for instance, be interpreted as an opportunity to give financial incentives to the part of the population negatively affected by the construction of wind and solar power.

### 3.3. Nuclear power option

Allowing for nuclear means expanding the available technology options, thus always inducing a decrease of System LCOE (or keeping the System LCOE at the same level as the base scenario). However, the resulting reduction of System LCOE is contingent on the costs of the alternative generation technologies, see Fig. 7. Two things stand out from the results: First, if the investment costs for PV and batteries are assumed to be low, allowing for nuclear power does not yield any reduction in System LCOE in MENA or Europe. Secondly, the cost reduction is smaller in MENA than it is in Europe. While the cost reduction in Europe is 11% and 19% for Mid and High investment costs for PV/batteries, respectively, the cost reduction in MENA is only a few percent, see Fig. 7.

The difference between MENA and Europe in the cost-reducing effect of allowing for nuclear (0–19% in Europe and 0–4% in MENA) may be explained by more favorable solar and wind resources in MENA. The abundance of solar and wind resources in MENA entails that a renewable system, including the necessary flexibility capacity (batteries and transmission), will out-compete nuclear in most subregions. In comparison, in Europe, there is less low-cost wind and solar resource in relation to its demand (see Fig. 1), which makes nuclear power relatively more competitive.

By modeling the power systems in MENA and Europe with and without the possibility of investing in nuclear power, we show that the impact from allowing for nuclear power on System LCOE may range between almost none to decrease System LCOE by about 20%. The effect of allowing for nuclear power is contingent on the supply of low-cost VRE resources (Fig. 1) as well as low-cost variation management resources (here battery storage). Thus, the higher-quality VRE resources in MENA, compared to Europe, entail that allowing nuclear power in the system has very little effect on System LCOE (the reduction in system cost is less than 4%, regardless of the cost assumptions for solar PV and batteries). In Europe, the benefit of including nuclear power is highly dependent on the cost assumptions, varying between 0% and 19% for the different cost assumptions for solar PV and battery investments. The System LCOE reductions that we find for Europe are in the low range of results in the literature, where e.g. Ref. [56] found a cost difference of between 11 and 44%, depending on the investment costs mainly for wind and solar. Reference [57] found only minor economic effects from allowing for nuclear in Switzerland. However, the literature also includes claims that decarbonizing without nuclear power (or other carbon-neutral thermal technologies such as coal wit CCS) is substantially more costly [17,55] than without those technologies. The main reason for these diverse conclusions in literature are likely different system boundaries, where regions are isolated and may not benefit from the flexibility provided by trade with other areas on a continental scale. In addition to this feature, which is also apparent in our study, we further nuance the potential cost benefit of nuclear to be contingent on resource quality and quantity in relationship to demand. Specifically, our study of MENA shows that the effect from allowing for nuclear power that was found in previous literature to be between 15% and 150%, depends on the regional characteristics of resource quality, availability of land and electricity demand. Thus, we argue that the results of comparative studies between $CO_2$ neutral power systems, such as [17,55–57]), should be interpreted bearing in mind the contingency both on method (isolated regions as in Ref. [17] or a larger connected region such as in Ref. [56]) as well as on the specific characteristics of the region, such as population density, electricity demand and availability of land for wind- and solar deployment.

Our results, which show that the reduction in System LCOE of allowing for nuclear power in MENA is very small, support strategies to decarbonize power systems in this region without investing in nuclear power and avoid associated concerns about safety and proliferation.

### 3.4. Transmission expansion

The present study investigated two scenarios regarding inter-subregional transmission capacity: an optimal expansion of the transmission capacity and a scenario excluding all transmission. We find an increase in System LCOE in both MENA and Europe when transmission is excluded, i.e., when subregions are isolated, see Fig. 8. The cost increase is similar in both regions: 4–23% in MENA and 11–25% in Europe, depending on cost assumptions for PV and batteries.

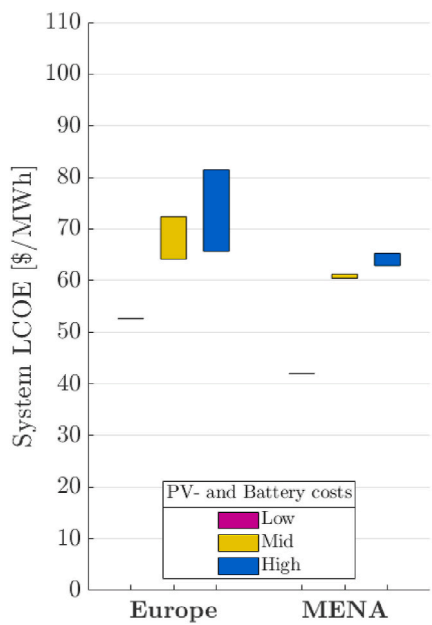

**Fig. 7.** Range of System LCOE between optimally installed nuclear power and no nuclear power for the cases of low, mid-range and high investment costs for PV and batteries.

---

[2] They formalize this concept by assuming a doubled investment cost for on-shore wind, thus reflecting increased cost for the entire building process.

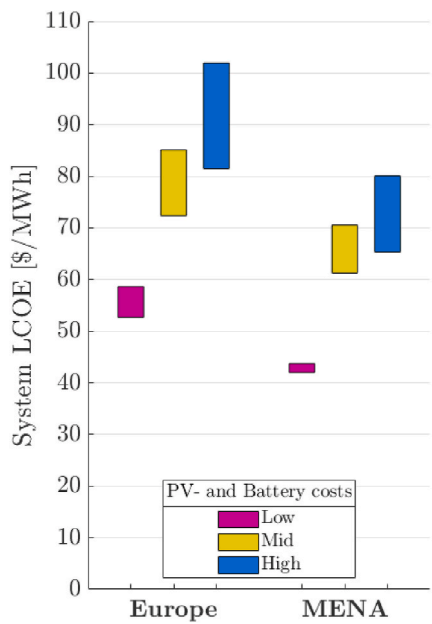

**Fig. 8.** Range of System LCOE between optimally installed transmission and no inter-subregional transmission for the cases of low, mid-range and high investment costs for PV and batteries.

Excluding inter-subregional transmission leads to a significantly higher System LCOE (23% and 25%, corresponding to a 15 $/MWh and 20 $/MWh cost increase respectively, for MENA and Europe), for high PV and battery investment costs, roughly corresponding to today's PV and battery costs. Conversely, the effect of excluding transmission is less significant when costs for PV and batteries are low. At low costs for PV and batteries, the PV-dominated system in MENA suffers only a small increase (4%) in System LCOE from excluding transmission (the corresponding number for Europe is 11%). The explanation is two-fold: First, allowing for transmission is more important when wind power is a large part of the mix. Hence, the smaller effect of excluding transmission when PV and batteries are low-cost and thus form a more substantial part of the mix, see Fig. 3. Second, if batteries are cheap, especially in combination with cheap solar, allowing for trade (through transmission expansions) has a smaller effect. The cost increase caused by excluding transmission also depends on the cost of the alternative technology, which here is wind power. The spatially scattered high quality wind power sites can not be used as efficiently without transmission as with transmission, causing the higher cost. Hence, the more favorable solar and wind conditions in MENA (see Fig. 1 and S3 - S7) is the reason for the slightly lower cost increase in MENA compared to Europe despite the higher share of wind power in Europe's energy mix when PV and battery costs are mid- and high-cost.

We find that System LCOE increases by between 4 and 25% when inter-subregional transmission is excluded. Isolating subregions leads to over-investment in VRE capacity and more investment in storage and thermal generation capacity with potentially low full-load hours. This cost increase is consistent with results from other modeling studies [6,7,10,11,13,14,58,59], where the cost decrease due to large-scale transmission expansion is found to be between 10 and 30%. However, unlike the majority of these papers, we investigate how the benefit of optimal transmission expansion depend on the cost of solar PV and battery storage. The benefit of adding inter-subregional transmission is greater when investment costs for PV and batteries are high compared to when they are low. A similar result was found by Schlachtberger et al. [4]. The underlying mechanism is that increased transmission mainly benefits systems with a high share of wind power, and low-cost solar PV and batteries lead to a smaller share of wind power in the optimal generation mix. Low-cost solar PV and batteries systems instead rely on more battery storage. For example, the low investment cost case results in 5.4 *TWh* of battery storage in MENA, equivalent to about 65 million Tesla model S (85kwh) batteries. This quantity of batteries may have consequences in terms of the use of materials. On the other hand, a large-scale grid extension might not be feasible due to social acceptance issues [35, 40–44] Thus, decarbonized power systems may entail hard-to-swallow features, such as large-scale transmission or large amounts of batteries or nuclear power.

*3.5. Limitations*

The main results of this paper are about how circumstances due to policy and public opinion (land for VRE, nuclear power, and the availability of transmission expansion) impact the cost of a $CO_2$-neutral power system, and how this impact may differ depending on regional resource endowment. The limitations of the model framework with potential consequences for this set of questions are:

1. The system boundary in this study is set around the electricity sector and does not include other sectors in the energy sector, such as heat, transportation, and industry. Besides, only one storage technology (lithium-ion batteries) is considered. This could have a bearing on the system cost difference between Europe and MENA. Sector coupling is likely to entail increased demand for electricity, both through electrification and the use of electro-fuels, thus increasing land scarcity for VRE farms. Thus, in the future, differences regarding the resource-to-demand relationship (Fig. 1) may be more considerable and have a more substantial effect on cost, thus increasing the cost in Europe compared to MENA. This would also impact the relative benefit of using nuclear for electricity generation, especially in Europe. Sector coupling increases the temporal flexibility of the system. In this sense, it resembles the effect of low-cost storage. Thus, if there is sector coupling, it is likely that the availability of other variation management strategies, such as transmission, becomes less consequential for System LCOE. Similarly, it would be comparatively less costly to integrate renewables, thus rendering the nuclear option less important for cost reduction. However, since sector coupling increases the demand for electricity, the land-availability issue becomes more pressing.
2. Political realities are not considered when modeling international transmission expansion, and only two cases are considered: no (international, i.e., inter-subregional) transmission and optimal transmission. These two extreme points of transmission expansion are both unlikely. In fact, transmission between subregions already exists in both Europe [90] and MENA [91]. The impact on System LCOE from extending transmission is in fact smaller than estimated in this paper, because the minimum amount of transmission is already greater than zero, and the maximum feasible transmission grid is likely smaller than the optimal grid.
3. We have not allowed for an expansion of hydropower. MENA exhibits a greater potential, compared to Europe, for expansion of hydropower. Allowing for expansion may therefore increase the difference in System LCOE between the regions.
4. We model every subregion as a copper plate, i.e., electricity transmission within each subregion is assumed to be unlimited. Due to this, internal transmission requirements are not considered. This assumption means that the cost for power systems is underestimated in general, and likely especially so in cases with large volumes of power traded between subregions. This model artifact could have a more significant effect on the cost estimate for MENA, since its model subregions are generally larger. This issue was addressed in Refs. [59], where it was seen that cost and capacity mix did not change significantly as the spatial resolution was gradually coarsened. However, it should be noted that the largest regional size in Ref. [59]

was still not as large as the regions in our study. The authors speculate that this is due to that, as the spatial becomes coarser, there are two mechanisms that counteract each other: the transmission needs are underestimated, but at the same time, the VRE resource is underestimated. This is due to that the method to estimate the VRE availability used in Ref. [59], entails that VRE resources are averaged, so that the best sites are no longer visible for larger regions. This latter is a trait which is less likely to interfere in the present study, since we employ wind- and solar classes in our model, thus capturing more of the resource heterogeneity compared with the method used in Refs. [59]. However, the lack of literature on the subject entails that we cannot be confident about the extent to which the large regions, and, especially, the unequal region size between Europe and MENA, impact the results. We believe that this topic merits more research in the future.
5. There are no ramping or start-and-stop costs for thermal power plants. Cebulla and Fichter [92] showed that including such constraints is of little consequence for predominantly renewable power systems.

We deem the first of these limitations as likely to have the largest effect on our results since it provides an alternative variation management strategy, which impacts the cost-effectiveness of nuclear as well as transmission expansion. Sector coupling also effectively increases the demand for electricity, putting more strain on land for power generation, which potentially increases the effect of less available land.

4. Conclusions

This paper investigates the effects of three socio-political factors on $CO_2$-neutral power system costs, the availability of: (i) nuclear power, (ii) international transmission, and (iii) land for wind and solar deployment. The analysis is applied to MENA and Europe separately, which allows for a comparison regarding how a priori conditions (such as population density, available land for RE and weather conditions) may be used to predict the cost and capacity mix of $CO_2$-neutral power systems. We find that:

- For any combination of assumptions on investment costs for solar PV and batteries, as well as transmission/nuclear/land availability, the system cost is lower in MENA than in Europe. This suggests that the lower system cost is linked to the better wind and solar resource quality.
- The cost for a $CO_2$-neutral power system ranges between 42 and 102 $/MWh in this study.
- Public acceptance of wind and solar farms may have a large impact on the cost of a $CO_2$-neutral power system. Our results indicate that a decrease of available land (from 10 to 2%) can increase system cost by about 50% in MENA and 25% in Europe for the case without nuclear but with the option of transmission expansion.
- Allowing for nuclear power reduces the system cost by 0–19% in Europe and 0–4% in MENA. The magnitude depends on investment costs for solar PV and batteries, resource quality, and the availability of land for wind and solar. Because these factors are more favorable in MENA, the availability of nuclear power has a greater impact on system cost in Europe than in MENA.
- Allowing for optimal transmission expansion decreases the system cost by between 5 and 25%. The highest cost decrease (25%) is found when PV and batteries are high-cost, due to a corresponding higher share of wind power which is favored by transmission, since it smooths out wind variations. The cost impact from optimal transmission is similar in Europe and MENA.

In summary, socio-political factors, here exemplified by whether nuclear power is included as a technology option, whether international transmission is possible, and the extent of land available for wind and solar farms, have markedly different impacts on results depending on the region (weather and demand conditions) and the cost of solar PV and storage capacity. Any judgment on the necessity of a specific socio-political factor for the realization of a decarbonized power system is contingent on assumptions regarding, for instance, investment costs and region. We also conclude that while the land available for wind and solar exploitation, which is affected by public acceptance issues, seems important for the system cost, this issue has not been investigated as thoroughly as other factors in model-based research. Future research could explore its importance in greater detail, with more realistic assumptions on restrictions for wind and solar expansion.

**Author contribution**

Hanna Ek Fälth: Conceptualization, Methodology, Formal analysis, Writing – original draft, Writing – review & editing. Dan Atsmon: Conceptualization, Methodology, Formal analysis, Writing – original draft. Lina Reichenberg: Conceptualization, Methodology, Writing – original draft, Writing – review & editing, Supervision. Vilhelm Verendel: Methodology, Writing – original draft

**Declaration of competing interest**

The authors declare that they have no known competing financial interests or personal relationships that could have appeared to influence the work reported in this paper.

**Acknowledgements**

The authors would like to acknowledge Niclas Mattsson for his generous help in using the GlobalEnergyGIS package.

**Appendix A. Supplementary data**

Supplementary data to this article can be found online at https://doi.org/10.1016/j.esr.2020.100590.